# Negative Results of Image Processing for Identifying Duplicate Questions on Stack Overflow

Faiz Ahmed
faiz5689@yorku.ca
York University
Toronto, ON, Canada

Suprakash Datta
datta@yorku.ca
York University
Toronto, ON, Canada

Maleknaz Nayebi
mnayebi@yorku.ca
York University
Toronto, ON, Canada

## ABSTRACT

In the rapidly evolving landscape of developer communities, Q&A platforms serve as crucial resources for crowdsourcing developers' knowledge. A notable trend is the increasing use of images to convey complex queries more effectively. However, the current state-of-the-art method of duplicate question detection has not kept pace with this shift, which predominantly concentrates on text-based analysis. Inspired by advancements in image processing and numerous studies in software engineering illustrating the promising future of image-based communication on social coding platforms, we delved into image-based techniques for identifying duplicate questions on Stack Overflow. When focusing solely on text analysis of Stack Overflow questions and omitting the use of images, our automated models overlook a significant aspect of the question. Previous research has demonstrated the complementary nature of images to text. To address this, we implemented two methods of image analysis: first, integrating the text from images into the question text, and second, evaluating the images based on their visual content using image captions. After a rigorous evaluation of our model, it became evident that the efficiency improvements achieved were relatively modest, approximately an average of 1%. This marginal enhancement falls short of what could be deemed a substantial impact. As an encouraging aspect, our work lays the foundation for easy replication and hypothesis validation, allowing future research to build upon our approach

## 1 INTRODUCTION

Community-driven question-answering platforms have gained traction for crowdsourcing information from software developers, thanks to the vast amount of data they accumulate through active user involvement. While various platforms cover a wide range of topics, such as Yahoo Answers and Quora, Stack Overflow stands out for its specific focus on software development. Since its inception in 2008, Stack Overflow has become a crucial platform for addressing programming issues. As of March 2022, Stack Overflow boasted over 20 million registered users, with over 24 million questions and 35 million answers [1].

To avoid bloating and better structure the forums, Stack Overflow encourages developers to thoroughly review existing posts before submitting a new question. This practice is designed to prevent the recurrence of queries that have already been raised and potentially resolved.[2] Despite these proactive measures, developers occasionally submit duplicate questions. In this context, duplicate questions refer to inquiries seeking solutions to identical problems. When two questions are identified as duplicates, one is flagged as such and undergoes the closure process, while the other is designated as the master. Experienced community members or users can cast votes to identify duplicates, and when a question garners enough of these votes, it will be closed as a duplicate [2]. Within the research community, to automate this process, some research efforts have been directed toward the identification of duplicate questions, emphasizing textual analysis and Q&A thread interactions [3–5]. Notably, an examination in a study conducted by Ahasanuzzaman et al. [6] showed that exact duplicates are infrequent; however, a multitude of questions are duplicated but phrased differently.

Fueled by advancements in Machine Learning, software engineers are increasingly interested in the role of images within developer communities, as evidenced by Nayebi et al. [7]. This has led to a surge of studies mirroring observations in social coding platforms, all highlighting a growing preference for image-based communication among developers. Platforms like Stack Overflow, a popular question-and-answer forum, have seen a rise in image usage, as documented by Nayebi et al. [7, 8] and Wang et al. [9].

On the other hand, a qualitative survey with software developers showed the complementary nature of these images to the textual information on bug repositories [7]. Often, the images are essential in understanding the change requests, questions, or responses submitted. Images help developers to convey complex information more effectively and efficiently than text alone. The study also pointed out that the majority of the images attached to questions were informative, and developers considered around 87.0% of the text unlikely or very unlikely to comprehend without the image attached to the question [7, 8]. They recognized that decision support systems that overlook image data are inherently biased, as they fail to consider the complementary nature of visual information. Furthermore, they anticipated that advancements in image processing and machine

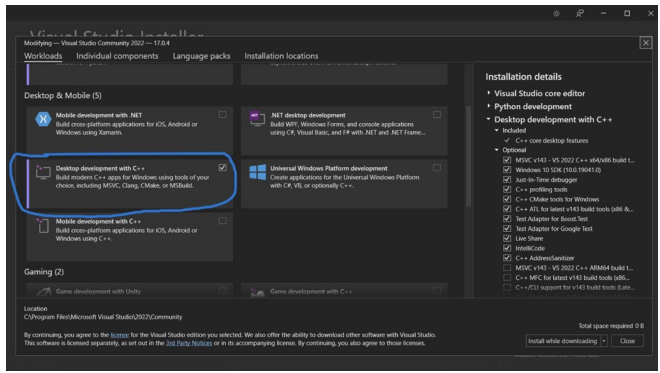

(a) Image attached to a question of our interest on Stack Overflow (Master Question)

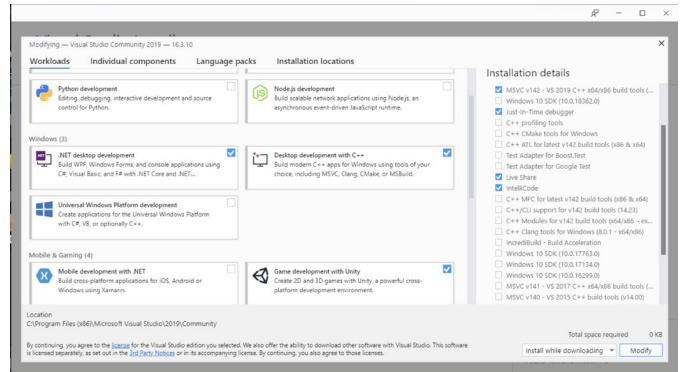

(b) Image attached to another question similar to our question of interest (Duplicate Question)

**Figure 1: Both the images are very similar even though they are attached to 2 different questions.**

learning would soon surpass the capabilities of systems that rely solely on text-based data on social coding platforms.

Drawing on the findings by Wang et al. [10] and Nayebi and Adams [8], which highlight the complementary nature of images shared alongside text, we propose that image processing can significantly enhance state-of-the-art duplicate detection. The reported 200% increase in the number of images shared on Stack Overflow [7] has motivated us to investigate whether the information in these images can lead to more accurate predictions of duplicate questions. Currently, automated methods for detecting duplicate content do not consider image information in their decision-making processes, prompting us to hypothesize that such decisions, based solely on partial information, are inherently flawed. Consider, for instance, Figure1, which displays two screenshots related to a bug fix. The visual similarities between these screenshots suggest that the associated questions might pertain to the same issue despite differences in their textual content.

Current text-based duplicate detection systems solely analyze question text, potentially overlooking valuable information embedded in attached images. This over-reliance on text can lead to biased and incomplete decisions, as crucial visual details might be missed that could be instrumental in determining question similarity. Our work aims to address this limitation by incorporating image data into the duplicate question identification process and assessing whether the current information systems lean toward biased decisions. By leveraging both textual and visual content, we aim to achieve more comprehensive and accurate duplicate detection on Stack Overflow. Specifically, we put forth the following research questions (RQs) for investigation:

**RQ1:** To what extent does the text extracted from images attached in Stack Overflow questions enhance the performance of state-of-the-art models for detecting duplicate questions?
**What and How:** The use of images in identifying redundant issue reports on Stack Overflow has the potential to enhance the accuracy of duplicate detection. Building upon past studies that indicate the complementary nature of images to text on Stack Overflow [7], we incorporate information from these images to evaluate whether it can enhance the performance of current methods for identifying duplicate Q&As. We use Optical Character Recognition (OCR) to this end [11, 12].

**RQ2:** Can incorporating the semantic meaning conveyed by image captions, alongside the existing focus on question text, significantly enhance the accuracy of duplicate question detection models on platforms like Stack Overflow?
**What and How:** We investigate how image information improves duplicate question detection in Stack Overflow. By generating captions for attached images using Google's Gemini-1.0-pro-vision model and incorporating them with textual features, we aim to create a richer representation for duplicate question identification. This approach leverages the complementary nature of image and text data to potentially enhance the accuracy of existing duplicate detection methods.[13]

**RQ3:** To what extent can combining image captions and Optical Character Recognition (OCR) for images in Stack Overflow questions improve the performance of duplicate question detection models?
**What and How:** Expanding on **RQ1** and **RQ2**, we investigate how image data combined with image text can improve duplicate question detection on Stack Overflow. We explore using image captions (via Gemini-1.0-pro-vision) and OCR (Nanonets) to extract text from images. Both extracted text and captions are incorporated with existing textual features. This evaluation helps us understand the effectiveness of image data for enhancing duplicate question detection accuracy.

Our study focuses on enhancing the current state-of-the-art in detecting duplicate questions on Stack Overflow by incorporating images, extending beyond existing methods that rely solely on text from the questions for duplicate detection. In this paper, we build upon the well-cited approach of Ahsanuzzaman et al. [6], which incorporates title, body description, tag similarity, and code to define and improve duplicate question detection. Our expansion involves incorporating image analysis to address **RQ1**, **RQ2**, and **RQ3**.

In what follows, we start by explaining our baseline study in Section 2. We then provide an overview of the empirical data in Section 3. Next, we introduce our research methodology, DupImage, for using images in identifying duplicates and answering our **RQs** in Section 4 and present the results in Section 5. We then discuss the importance of publishing negative results in Section 6 and discuss the related works in Section 7. Lastly, we conclude by discussing the threats to validity and providing a conclusion (Section 8 & 9) and providing the details of our replication package in Section 10.

## 2 BASELINE FOR IDENTIFYING DUPLICATE QUESTIONS

In 2015, Zhang et al. [4] introduced a model, DupPredictor, for identifying duplicate questions in StackOverflow. Later in 2016, Ahsanuzzaman et al. [6] further extended this model into Dupe. For answering our research questions, we employed the Dupe model developed by Ahasanuzzaman et al. [6] as the extensively cited, frequently reused [3, 14], and replicated approach [5]. This choice served as the basis for our evaluation, where we investigated the potential enhancement of its performance through image processing in the identification of duplicate questions on Stack Overflow.

The approaches presented by DupPredictor [4] and Dupe [6] were evaluated on a dataset containing Stack Overflow questions up to September 2014 [15]. This dataset included 130,888 non-master duplicates and 90,245 master questions. We followed the state-of-the-art methods [16–26]. A variety of analytical models for software and project management exists in software engineering research [27–33] which we build our study upon such studies [7, 8, 19, 24, 34–49].

### 2.1 DupPredictor Methodology

Zhang et al. [4] proposed an approach to address the problem of solving duplicate question detection on Stack Overflow. This approach takes a new question as input and returns a ranked set of questions that potentially have high similarity with the input. The DupPredictor framework comprised two primary phases: (i) model-building and (ii) predicting duplicates.

In Phase (i), historical questions were gathered from Stack Overflow to serve as training data. Step 2 focused on preprocessing, involving the extraction of title, description, and tags from each question, followed by text tokenization, removal of common English stop words, and stemming using the Porter stemming algorithm [50]. After looking at different parts of questions (title, description, topic, and tags) explained below, the system calculated a score for how similar each question was to the original question. These scores were then combined into a single overall similarity score. This final score was used to rank other questions based on how likely they were to be duplicates of the original question.

In their approach, Zhang et al. [4] measured similarity across several aspects of questions to identify duplicates. Title similarity considered how many common words question titles shared, leveraging a topic model [51]. Description similarity focused on the overlap of words within the question descriptions after converting them to a simplified format for comparison. Finally, tag similarity assessed how closely related the tags assigned to each question were, providing another indicator of how well the questions might address similar topics [4].

Once the model is trained in Phase (i), we transition to Phase (ii), where the trained model is applied to detect new duplicate questions by comparing each new question with all past questions on Stack Overflow and computing similarity scores. For measuring similarity, similar to Phase (i), new Similarity scores are calculated between the new question and each historical question using the same four factors (title, description, topics, and tags). After calculating each similarity, a final consolidated similarity score is calculated to rank the historical questions. Questions with higher scores are placed higher in the ranked list, indicating a greater likelihood of being duplicates of the new question. As a result, the model outputs the top-K most similar questions, providing a curated set of historical questions that potentially mirror the new question.

### 2.2 Dupe Methodology

Expanding upon the DupPredictor, Ahasanuzzaman et al. [6] introduced the Dupe method to address the issue of duplicate questions on Stack Overflow. The Dupe and DupPredictor methods share the common goal of identifying potential duplicates based on title, textual content, topic, and tag similarity. However, they diverge in several aspects. Firstly, Dupe employs a discriminative classification model [52] for duplicate detection, while DupPredictor relies on a combined similarity score. Secondly, Dupe incorporates a more extensive set of features compared to DupPredictor. Additionally, Dupe looks into the characteristics of duplicate questions and investigates the reasons behind their occurrence, aspects not covered in DupPredictor's methodology. Dupe doesn't just find these duplicates, it also tries to understand why they happen in the first place. It does this by looking at the way the questions are written, like the words they use and how they're phrased. This might reveal reasons like people not searching well or asking questions in a confusing way [6]. By figuring out these reasons, Dupe aims to make Stack Overflow a better experience for everyone. It can potentially help reduce the number of duplicate questions and make searching for answers easier.

Dupe operates through three main phases. In the initial phase, the text of the questions undergoes preprocessing. The second phase involves collecting various features, as discussed below, for each question pair (where a pair consists of a question and its duplicate) to train a binary classification model. This model plays a crucial role in the third phase, where duplicate question detection takes place.

In the initial phase of Dupe, the text undergoes preprocessing which involves removing the stop word, performing stemming, and converting tags into a unified form. On Stack Overflow, tags follow a master-synonym structure. A tag synonym is a tag that shares an identical meaning with another tag. In some cases, a tag can be a subset of another tag, and this is also deemed synonymous. They substitute each instance of a synonym tag with its corresponding master tag to enhance the duplicate detection process

The second phase of Dupe employed a binary classifier to pinpoint duplicate questions. This classifier relied on five key features. Cosine similarity, based on the vector space model, evaluated the similarity between questions by considering the angles between their vector representations. Term overlap directly measured how many identical words the questions shared. Entity overlap, measured using the Jaccard coefficient, captured the proportion of shared entities (like names or keywords) between the questions. Entity type overlap went beyond just the entities themselves, considering if the types of entities also overlapped. Finally, WordNet similarity, leveraging a Java library (WS4J), calculated semantic relatedness between questions using the WordNet knowledge base, identifying deeper conceptual similarities beyond just surface-level keywords.

Ahasanuzzaman et al. [6] analysis in Dupe by drawing insights from their results showed that cosine similarity is the most valuable feature in this context, while other features provided little to no improvement in the model's performance. With these features, for a given pair of questions and their corresponding feature values, the classifier determines whether they are duplicates and assigns a probabilistic value indicating the level of duplication. Given the potentially large number of question pairs classified as duplicates, relying solely on a binary classifier is insufficient. Therefore, in the third phase, similar to DupPredictor, they output the top k probable duplicates by sorting the questions based on their duplication levels to generate a ranked list of results. This ranking is determined by the likelihood of the new question being a duplicate of another in the Stack Overflow repository.

## 2.3 Evaluation Metrics

The promising results of these models were further extended by multiple researchers in a series of models [5] [3] [14]. In all these studies, including DupPredictor and Dupe, to evaluate the performance of the proposed techniques, the authors used the recall rate.

$$recall-rate_k = \frac{N_{dk}}{N_{all}}$$

Here, $recall-rate_k$ refers to the recall of a technique within top-k recommendations. The term $N_{all}$ refers to the total number of duplicates in our test data and the term $N_{dk}$ refers to the total number of detected duplicates in the top-k recommendations. In the previous works [6] [5] [4], k is set to 5, 10, and 20, respectively, and we also evaluate our results based on these three metrics.

The best results achieved by both models are presented in Table 1. In the baseline studies, authors divided the dataset into different question groups such as Java, Python, Ruby, C++, HTML, and Objective-C. As the table clearly shows, Dupe outperformed DupPredictor in terms of performance.

## 3 EMPIRICAL DATA

This section details the processes involved in constructing our experimental dataset. We focused on gathering duplicate and non-duplicate question-image pairs to investigate the role of visual information in duplicate question detection. We discuss the process we applied for data retrieval and creation in this section.

We use this data to evaluate baselines explained in the previous section and perform benchmarks. We also use this data to evaluate whether the use of images would increase the chances of identifying duplicate questions on Stack Overflow and answering our RQs.

**Table 1: Baseline Model Results for DupPredictor and Dupe**

| Baseline Model | Question Group | Best Recall Rate | |
|---|---|---|---|
| DupPredictor [4] | Ruby | Top 5 (%) | 33.00 |
| | | Top 10 (%) | 37.00 |
| | | Top 20 (%) | 39.00 |
| Dupe [6] | Ruby | Top 5 (%) | 51.16 |
| | | Top 10 (%) | 59.56 |
| | | Top 20 (%) | 66.11 |

## 3.1 Data Gathering Process

For our study, we sourced data from Stack Overflow by querying the Stack Exchange Data dump [53]. The data spans several years since the inception of Stack Overflow in 2008 up to December 31$^{st}$, 2022. We had two criteria for including images in our set:

**Criteria 1- Questions marked as duplicate:** Our focus was on identifying duplicate questions within the Stack Overflow platform. When a question is identified as a duplicate, it is closed and linked to a "master question" containing the solution. This redirection ensures users can efficiently access relevant information. To achieve this, we adopted the approach established by Ahasanuzzaman et al. [6]. Their work identified a specific marker used to flag duplicate questions. We leveraged the Stack Exchange Data Dump [53] to extract these questions using targeted queries. To clearly distinguish them within our dataset, we appended the string "[Duplicate]" to the titles of these questions.

**Criteria 2- Questions with an image:** Given our primary focus on the role of images in question duplication, we specifically tailored our query to filter for questions with an attached image in the body section of the posts. We can easily get only those posts which contain these images by integrating a condition to look for `` tag in the body section of each post when querying from the Stack Exchange Data Dump [53]. This targeted selection strategy helps us in getting only those posts which have images attached to them.

By querying the *posts database* in the Stack Exchange Data Dump along with these two criteria, we were able to extract 42,462 non-master duplicate questions. To form a fair and balanced baseline for comparison, we randomly queried a total of 20,493 non-duplicate questions that were closed and marked as non-duplicates and were not marked as masters of any other question.

Non-master duplicate questions are those identified as duplicates of other questions and, as a result, were closed without extensive discussion. These questions are linked to a master question. Following this, we sought to identify the 'master' questions to which these duplicates were linked. Master questions serve as the original or primary questions that the duplicates are referencing. To collect these master questions, we queried the `postlinks` and `posts` databases in the Stack Exchange Data Dump, identifying a total of 6,265 master questions [6].

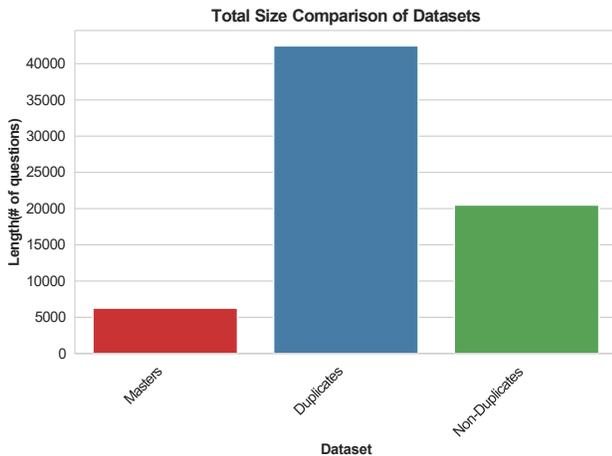

**Figure 2: Overview of the size of the dataset**

It's important to note that all the questions in our dataset, the duplicates, non-duplicates, and the master questions, have images attached to them. This unique characteristic of our dataset will allow us to explore the role of images in detecting duplicate questions with greater specificity and accuracy.

### 3.2 Data Set Creation

After extracting each question from the data gathered in Section 3.1 we applied a series of preprocessing steps. We followed the process explained in the baseline model Dupe [6] for this purpose. Firstly, we separately collect the title, body, tags, and image URLs for each question from the data gathered in Section 3.1 using `BeautifulSoup` library. Then, for each question in the gathered data, we remove stop words from the title and body. Next, to prevent unwarranted discrepancies, we apply stemming to textual features (title and body content) using Porter's algorithm following the methodology in Dupe [6]. Stemming simplifies words by removing common endings. This helps make sure all similar words are treated the same way in our analysis.

Following the methodology in Dupe by Ahsanuzzaman et. al. [6], we subsitute each instance of a synonym tag with its corresponding master tag to enhance the duplicate detection process. In the subsequent phase, we establish connections between duplicate questions and their corresponding master questions, thereby forming pairs. For every duplicate question, we examine the *RelatedPostId* to locate its master question. Provided that the master question is present in the data compiled in Section 3.1, we generate a pair for that particular question. This process resulted in a total of 6,265 pairs of duplicate questions.

Following a similar process to the previous step, we generate non-duplicate question pairs. This is achieved by randomly associating master questions with 6,265 non-duplicate questions randomly selected from the 20,493 non-duplicate questions gathered in Section 3.1, thereby creating an equivalent number of non-duplicate question pairs. We can be sure that they are not duplicates because non-duplicate questions are closed and marked as not duplicates of any other question in stack overflow.

We then filter out those questions for which the image URL was inaccessible. This was necessary to ensure the integrity and usability of our data. Given the resource-intensive nature of image processing tasks, we decided to reduce the complexity of our problem. We achieved this by dividing our dataset in half which resulted in 3,003 pairs of duplicate and non-duplicate questions, which served as the dataset for our model. This selection strategy acknowledges resource constraints and the time required for image processing, while still providing a sufficient data volume for effective model training.

***Note:*** *All the questions in the comprised dataset at the end of Section 3 have images attached to them.*

## 4 PROPOSED METHODOLOGY: IMAGE-BASED DUPLICATE DETECTION (DUPIMAGE)

We started by replicating Dupe. To identify duplicate questions, we adopt the same methodology used in Dupe [6], which involves the application of logistic regression to create our classification model. Logistic Regression is a type of classifier that works with real-valued vector inputs. It's a probabilistic classifier, meaning it generates a predictive probability for each class, indicating the likelihood of a test case belonging to that class [54]. We focus on these probability values rather than binary classification, as our goal is to identify the top-k questions and determine if any duplicates exist within this subset. To this end, we use the data set created in the Section 3.2.

To incorporate images into the detection process, we identified three ways for integrating data by either using the textual data from the images (**RQ1**), or using image captions (**RQ2**). We also combined the two by integrating textual content in the images with the captions generated from the question to form a more comprehensive representation of the issue (**RQ3**). This image text, image captions, and combined data can then be input into a machine-learning model (logistic regression in our case) to identify duplicates based on both textual content and image captions. We used these in addition to the Dupe model [6] to evaluate whether this improve its performance. We can extract relevant data from the images using techniques such as Optical Character Recognition (OCR) and the image captions generated using the gemini-1.0-pro-vision model. OCR can be used to extract any text present in the image, while image captions can be used to capture a broader visual understanding of the images.

### 4.1 RQ1: Mining text from image with Optical Character Recognition (OCR)

Since we are focusing on finding duplicate questions on Stack Overflow, the images we will be looking at are most likely screenshots, bits of code, diagrams, charts, or user interface mockups (UIs). Because questions with similar problems are likely to have similar text in their images, it is crucial to be able to read the text within these images. This is where Optical Character Recognition (OCR) proves valuable and helps us identify the text in the images.

OCR is a computer vision technology that primarily aims to convert non-editable textual content embedded in images, scanned documents, or other visual formats into machine-readable text [11, 55, 56]. The process involves these main steps: First, the image might be converted to grayscale and cleaned up to make the text

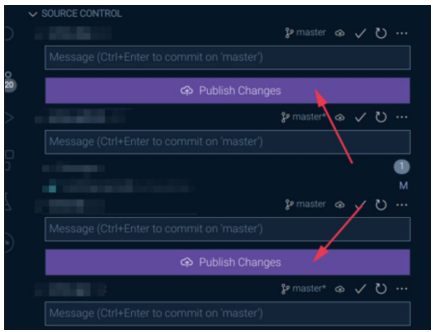

**Image Caption:**
Click the **Publish Changes** button to publish your changes.

**OCR Text:**
SOURCE CONTROL I f master Message Ctr+Enter to commit on master C Publish Changes Ps master Message Ctr+Enter to commit on master i M f master Message Ctrl+ Enter to commit on master O Publish Changes is master a Message Ctrl+Enter to commit on master I

**(a) Image Captions and OCR for an image in Master Question**

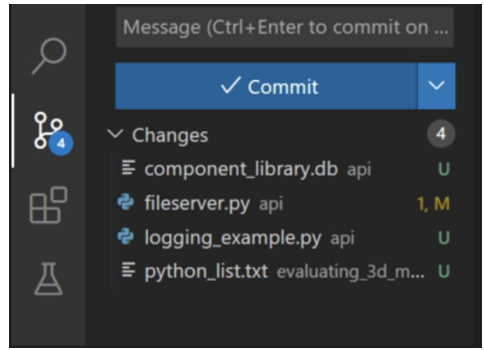

**Image Caption:**
The image shows a screenshot of a Visual Studio Code window. The window is split into two panes. The left pane shows the source control changes. The right pane shows the files that have been changed.

**OCR Text:**
( Message (Ctrl+ Enter to commit on . O VCommit 8% V Changes 4 a component library.db api U R fileserver.py api 1. logging_ example.py api U a a python list.txt evaluating 3d m .. U

**(b) Image Captions and OCR for an image in Duplicate Question**

**Figure 3: Both the images are talking about committing changes and the text and captions generated are very relevant. OCR Text indentation has been removed to make it look cleaner.**

clearer for the computer to read. Then, the image is divided into smaller chunks, separating individual letters from the background. From each chunk, features like shape or edges are extracted. Finally, these features are compared to a database of known characters (like letters and numbers) using clever algorithms to determine the most likely character in each chunk. By combining these steps, OCR systems can essentially create a function that takes an image region and outputs the most likely recognized character based on what it "sees" in the image. [12]

For reference, take a look at an image attached to a Stack Overflow question where a user is encountering issues with version control in Visual Studio Code (VSCode) in Figure 3. The screenshot might depict the changes staged for commit within the VSCode interface. OCR can take screenshots like this and convert the text within the image, such as file names or error messages, into a format the computer can understand. By integrating the extracted text from the image along with the written text of the question itself, it seems that we can improve our chances of finding duplicates. Textual similarity of images becomes another data point for identifying questions that might be addressing the same issue, even if the phrasing in the written text differs slightly.

For the purpose of our study, we used Nanonets [57] to implement OCR in our DupImage. Nanonets is a deep learning-based library with a high-accuracy implementation for the OCR task. The model is based on advanced convolutional and recurrent neural networks. For our research paper, we have chosen to utilize Nanonets' OCR technology, as it outperforms all other existing Character Recognition technologies in terms of accuracy and efficiency [57].

## 4.2 RQ2: Image Captioning with Gemini-1.0-Pro-Vision Model

In the previous section (Section 4.1), we discussed using Optical Character Recognition (OCR) to read text from images. As evident, it just focuses on extracting the text from the images and no semantic understanding of the images is captured. To address this issue, we incorporated image captioning as it helps capture the semantic meaning of the images. We utilized a powerful Large Language Model (LLM) from Google called Gemini-1.0-Pro-Vision to automatically describe the visual content within images attached to Stack Overflow questions. These descriptions can be very helpful for finding duplicate questions, especially when the image itself might not be very informative or does not contain text for OCR. To make it more robust, we performed image captioning on each question pair 3 times and chose the captions that gave us the best similarity scores. This was done to avoid hallucinations by the LLM model as it can generate different captions for the same image.

To understand the image captioning better, take a look at the images attached to questions about version control in Visual Studio Code (VSCode) that includes a screenshot in Figure 3. While the screenshot might not explicitly show the VSCode icon or other branding elements, image captioning can still be highly effective and identify these IDEs which can help us better compare the images. This technique can analyze the overall layout of the screenshot, the presence of buttons or menus, and potentially even text snippets. By generating a description that mentions meaningful elements from the images like "source control changes" or "code editors", image captioning can provide valuable insights. This additional information, beyond just the extracted text from OCR, can significantly improve our ability to identify duplicate questions that might be addressing the same underlying issue.

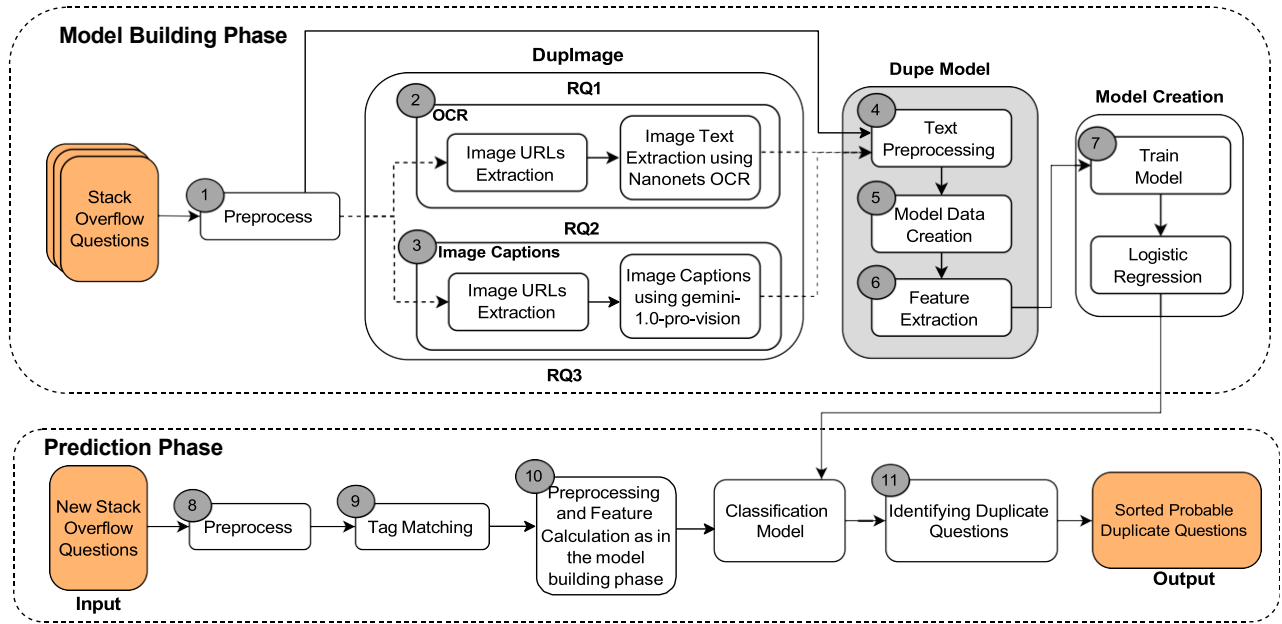

**Figure 4: Overview of the processes used for detecting duplicate questions on Stack Overflow: Text, OCR, and Captions**

Compared to Optical Character Recognition (OCR), which excels at extracting text embedded within Stack Overflow question images (code snippets, error messages, etc.), image captioning offers a broader perspective on the visual content. While OCR focuses on the "what" (written text), image captioning can draw insights from the "why" and "how" aspects of the image. It can capture objects, actions, and relationships depicted in the image that might not be explicitly present as text.

Google's `gemini-1.0-pro-vision` model leverages the power of deep learning, specifically convolutional neural networks (CNNs), to analyze images and generate captions that capture their essence [58, 59]. By going beyond basic description, these captions provide a richer understanding of the image's content.

### 4.3 RQ3: Combination of Optical Character Recognition and Image Captioning

Building upon the individual strengths of OCR (Section 4.1) and image captioning (Section 4.2), RQ3 explores how their combined power can enhance duplicate question detection on Stack Overflow. OCR excels at extracting text from images, providing valuable keywords and phrases that can directly relate to the problem at hand (Section 4.1). Image captioning, on the other hand, goes beyond just text by analyzing the visual content and generating descriptions that capture the overall context of the image (Section 4.2).

In our work, we aimed to develop a robust similarity measure for duplicate question detection on Stack Overflow. This measure leverages the strengths of both textual and visual information embedded within questions. We combined similarity scores obtained from two sources: 1) Optical Character Recognition (OCR) to extract text from question images (**RQ1**), and 2) image captioning using `gemini-1.0-pro-vision` to generate captions describing the visual content of the images (**RQ2**). To achieve a more comprehensive similarity score, we employed a max function, selecting the higher score between the image text similarity and the image caption similarity for each question pair. This approach addressed scenarios where images might have imbalanced information distribution. For instance, two images might have different text but similar visual content which would create similar captions, resulting in a higher image caption similarity score. For example, if images of graphs are attached talking about the population increase without much image text but with high semantic meaning. By considering the maximum similarity score, our method could effectively capture these variations and identify potential duplicates.

$$S_{\text{combined}}(q_i, q_j) = \max(IT_{\text{sim}}(q_i, q_j), IC_{\text{sim}}(q_i, q_j)) \quad (1)$$

where

- $S_{\text{combined}}(q_i, q_j)$ is the combined similarity score between question $i$ and question $j$.
- $IT_{\text{sim}}(q_i, q_j)$ is the image text similarity score between question $i$ and question $j$, obtained from OCR in Section 4.1.
- $IC_{\text{sim}}(q_i, q_j)$ is the image caption similarity score between question $i$ and question $j$, obtained from image captioning in Section 4.2.

To confirm the effectiveness of our approach (using the max/higher similarity score), we manually analyzed some question pairs. We calculated a value called "delta" for each pair. This delta simply represents the absolute difference between the two similarity scores (image text and image captions). Question pairs with a significant delta indicated a big difference in scores. In our case, for manual

Table 2: Evaluation of "recall rates" to measure the impact of different Image Comparison Techniques

| No. | Configuration | Top 5 (%) | Top 10 (%) | Top 20 (%) |
|---|---|---|---|---|
| 1 | Replication Results on Image Dataset (Dupe) | 43.43 | 54.24 | 61.06 |
| 2 | Only OCR | 21.63 | 25.12 | 30.28 |
| 3 | Only Image Captions | 13.98 | 16.98 | 24.28 |
| 4 | OCR with Dupe Text Model | 45.42 | 55.57 | 62.23 |
| 5 | Image Captions with Dupe Text Model | 43.93 | 54.25 | 61.23 |
| 6 | OCR + Image Captions + Dupe Text Model | *45.26\** | 55.24 | *62.37\** |

analysis, we considered a question pair to have a significant difference if the delta was higher than 0.5. This threshold gave us a total of 485 pairs with a 'delta' value greater than 0.5. We then closely examined these question pairs with a large difference in scores. Even though the image text and image caption similarity scores themselves might not have been very close, in most cases, we found that the questions were duplicates. This reinforces the importance of using information from both the image text and the visual content of the images. Some images might have very little written text, but the image might clearly show what's going on. By considering both text and semantics, we can improve our ability to identify duplicates, even when the information is presented in different ways

$$\delta(q_i, q_j) = |IT_{\text{sim}}(q_i, q_j) - IC_{\text{sim}}(q_i, q_j)| \quad (2)$$

where

- $\delta(q_i, q_j)$ is the delta value representing the absolute difference between image text and image caption similarity scores for question $i$ and question $j$.

## 5 EVALUATION RESULTS

In this section, we present the results of our benchmarking to answer the **Research Questions** in the designed case study and illustrate the experimental results of our approaches compared to the baseline approach of Dupe. We run all our experiments on the *image dataset* which comprises 3,003 non-master duplicate questions, 3,003 non-duplicate questions, and 3,003 master questions as discussed in Section 3. We use this dataset with images in the questions as the main focus of the paper is to identify how important images can prove to be in the context of detecting duplicate questions on Stack Overflow. Moreover, we have not divided our dataset based on the question groups as in the previous approaches [6] [4] [5], as the size of our image dataset is reasonable enough.

We allocate 80% of both the duplicate pairs and non-duplicate pairs datasets, discussed in Section 3.2, for training our model, ensuring a balanced representation to prevent any training bias. The remaining 20% of the duplicate-pairs dataset is used for model evaluation. Based on previous methodologies [5, 6], where they used cosine similarity value as a significant feature for detecting duplicate questions, we also incorporate cosine similarity as a feature in our baseline approach, applying it to title-title, title-body, body-body, and code-code comparisons.

Firstly, we integrated the image text obtained through the OCR technique, as outlined in Section 4.1, and computed the cosine similarity between vectors of image text pairs as part of our model evaluation. Next, we employed the image captioning method to determine the similarity score between images based on key visual descriptions, as detailed in Section 4.2. Lastly, we looked at the potential of combining these two techniques and assessed their collective impact on detecting duplicate questions on Stack Overflow as discussed in Section 4.3.

### 5.1 RQ1: Results of integrating OCR

In our research, to answer **RQ1**, we wanted to see if using images could help find duplicate questions on websites like Stack Overflow. We realized that images in these questions often contain important text that explains the problem. So, we developed a method (explained in Section 4.1) to find and analyze this text from the images. Our goal was to find similarities between the text in the images and use that to identify duplicate questions. By treating the text in images as valuable information, we aimed to connect the traditional text-based analysis with the image-heavy world of online Q&A platforms.

Unfortunately, our findings as depicted in Table 2 show that extracting text from images for comparison didn't significantly improve our ability to find duplicate questions which is measured by $recall-rate_k$. While using Optical Character Recognition (OCR) alone performed poorly, even combining it with the existing method of Dupe only led to a small improvement (around 2% on average). These results suggest that focusing on text analysis within images wasn't as effective as we hoped for duplicate question detection.

One potential limitation of using OCR for duplicate question detection is the inherent similarity between images in related questions. Even when questions share similar meaning or intent, the actual text displayed within the images might differ. For instance, questions about troubleshooting a specific software issue could both depict screenshots of the program's interface, yet the specific error messages or menu options displayed might vary. This text discrepancy can get in the way of the effectiveness of OCR in accurately identifying duplicate questions. Additionally, OCR might struggle with poorly lit or blurry images, further reducing its accuracy in capturing relevant textual information. However, Nanonets' OCR is pretty robust in detecting text on dark backgrounds.

### 5.2 RQ2: Results of integrating Image Captions

In our study, for RQ2, we looked at using image captions to understand the semantic meaning of images attached to questions on Stack Overflow. These captions can be very helpful, especially when questions include similar pictures with minimal text in them. We developed a method (explained in Section 4.2) to automatically generate captions describing what the images show. Our goal was to find duplicate questions by comparing these captions.

However, the results we obtained, as illustrated in Table 2, were not up to the mark. While we anticipated that extracting conceptual information from images through captions would improve duplicate question detection, the results did not demonstrate a significant increase in the $recall-rate_k$. Image captioning on its own, as our results in Table 2 showed, did not perform as well as we had hoped. Even when we combined this approach with the existing method of Dupe, the improvement in recall rate was still relatively small, typically around 1% on average. These findings suggest that focusing on understanding the content described in image captions wasn't as successful as we had hoped for identifying duplicate questions.

While image captioning holds promise for identifying duplicate questions, its effectiveness can be limited by the capabilities of Large Language Models (LLMs) used to generate captions (`gemini-1.0-pro-vision` in our case). LLMs can sometimes create inaccurate captions ("hallucinate") or focus on irrelevant details in images. For example, similar error messages on different operating systems like Linux and Windows might be interpreted differently by the LLM, leading to dissimilar captions and predicting duplicate detection wrong. Recognizing these limitations is crucial for refining image captioning techniques for this specific task.

### 5.3 RQ3: Results of the combination of OCR and Image Captions

In our study, to answer **RQ3**, we also investigated the potential of combining OCR for text-based analysis and Image Captions for image content analysis to enhance duplicate question detection in the Q&A platforms. While images play a significant role in Q&A platforms, the integration of these approaches, even when combined, did not result in a substantial improvement in the $recall-rate_k$, as illustrated in Table 2. While our investigation yielded some improvement in duplicate question detection, the overall improvement was not significant enough to be able to consider this as a successful integration.

Our research aimed to enhance the performance of existing state-of-the-art models, such as Dupe [6], for duplicate question detection on Stack Overflow by incorporating information from question images. While we successfully integrated image data, a substantial improvement in overall performance was not achieved. This finding suggests that incorporating visual information, while a promising avenue for future research, may require more sophisticated techniques to unlock its full potential for duplicate question detection.

Although the improvement achieved through the combination of OCR and Image Captioning is marginal, on average of around 2%. The findings suggest that images do provide supplemental information to the text within the question (title, body, and code). However, we categorize this result as negative due to the increased complexity introduced by integrating these image-processing techniques. Generating OCR text and image captions for the entire duplicate questions with images dataset (DupImage dataset) required approximately 12 hours each on Apple M2 Pro with 10 cores. This significant processing time highlights the computational burden associated with these methods.

## 6 IMPORTANCE OF NEGATIVE RESULTS

The call for ESEM 2024 encouraged authors to publish negative results. While we slightly could demonstrate an improvement of 2% when using images compared to the pure analysis of text with methods such as Dupe, we consider this improvement trivial, which does not justify the allocated resources and the complexity involved in image processing.

Discussing negative results in empirical software engineering is crucial for several reasons [60, 61]. First and foremost, sharing negative results helps to avoid duplication of effort within the community. By clearly demonstrating that integrating image processing for our specific application resulted in only a marginal improvement, other researchers can make more informed decisions about resource allocation. This further is complemented by the comprehensive replication package of not only our DupImage method and data set but also the replication package we created for Dupe. Furthermore, publishing negative results contributes to a more comprehensive understanding of the problem space. It helps to map out the boundaries of effective techniques and can guide future research toward more promising areas or novel approaches [62, 63]. Additionally, acknowledging and discussing negative results enhances the credibility and authenticity of the research process. It reflects a commitment to transparency and rigor, rather than merely pursuing and highlighting positive outcomes.

## 7 RELATED WORK

We discuss the problem of identifying duplications in software repositories and the use of image processing in software engineering in this section.

### 7.1 Identifying duplicate entries in Software Engineering

Identifying duplicate bug reports in issue-tracking systems is crucial for software engineers to enhance software teams' efficiency and software developers' productivity. Sun et al. [64] used discriminative models for information retrieval and achieved substantial 17–43% relative improvements in OpenOffice, Firefox, and Eclipse datasets, relying solely on common natural language information.

Sometimes these duplicate bug issues can prove to be useful, Bettenburg et al. [65] conducted a survey and found that while most developers have experienced duplicated bug reports, only a few considered them as a serious problem. Meanwhile, a number of studies have developed methods to identify duplicate bug reports [66–70]. Similarly, with the emergence of crowdsourcing Q&A platforms like Stack Overflow, identifying duplicate questions is a research focus. Studies showed that the majority of duplicated questions are re-asked only once, approximately 15% are posed more than twice, with some reaching up to 558 times, resembling frequently asked questions rather than typical duplicates [71]. However, some argued the positive impact of duplicate questions on developers' understandings [72]. Apart from DupPredictor [4] and its extended method Dupe [6], as well as the replication of that by Silva et al. [5], which we employed as the baseline in this study, several other studies have explored the application of deep learning for enhanced text analysis to identify duplicate questions. Wang et al. [3] utilized deep learning techniques, such as Convolutional Neural Networks (CNN), Recurrent Neural Networks (RNN), and Long Short-Term Memory (LSTM), to detect identical questions on Stack Overflow. Additionally, Liao et al. [14] employed the Siamese BiLSTM method to capture semantic and contextual relevance between questions, addressing the duplicate issue. They reported a moderate improvement compared to the previous deep learning model.

## 7.2 Use of Images for Software Engineering tasks

In recent years, the importance of visual content in software engineering has been increasingly recognized. The rise of social coding environments and social networks has led to a surge in the sharing of images among developers [7]. Recent studies on Bugzilla and Stack Overflow over nine years reveal a doubling in image-inclusive posts between 2013 and 2022. These images, often essential for understanding the content, increased engagement from other developers. In 86.9% of cases, comprehension without the image was unlikely, emphasizing the need for visual content consideration in software engineering tool design and developer behavior analysis. [8]. Automated testing for cross-browser compatibility often struggles with false positives. Browser Bite, a tool leveraging image segmentation, differencing, and machine learning, demonstrate significant improvement, achieving an F-score exceeding 90%. This highlights the potential of image-based machine learning for more accurate software testing [73]. TANGO, a technique for identifying duplicate bug reports from videos using computer vision and text analysis shows that TANGO can effectively rank duplicates and significantly reduce developer effort compared to manual analysis [74].

A recent study [75] proposed 23 image features for classifying UML class diagrams (UML CDs). It found 19 of these features to be influential predictors, with six classification algorithms achieving a prediction rate of nearly 96% for UML CDs and 91% for non-UML CDs. Moreover, another study [10] reveals that questions with images are more likely to get accepted answers than those without images on platforms like Stack Overflow, demonstrating the importance of images. For instance, Calefato et al. [76] discovered a positive correlation between including code snippets and the success of questions. Jason et. al. introduced WebUI, a large dataset of automatically collected web pages with extracted visual information. The authors explore using web semantics to improve visual UI understanding models, especially for mobile applications with limited labeled data [77].

GIFdroid, a lightweight approach using image processing to automatically replay bug reports from screen recordings by generating the execution trace [78]. Existing GUI grouping methods rely on supervised learning or heuristics, while this work proposes a novel unsupervised image-based approach inspired by perceptual grouping principles [79]. However, the influence of images, a commonly used non-textual element in today's society, is still largely unexplored. Therefore, we aim to investigate the potential of using images accompanying questions to detect duplicate queries, a task of significant importance.

## 8 THREATS TO VALIDITY

Just like any empirical study, ours has some limitations that we need to consider. These limitations can affect how well our findings apply to other situations.

**External Validity:** The generalizability of our findings depends on the representativeness of our chosen sample. The types of questions we looked at and the time period we covered could influence how well our conclusions can be applied to other situations. Additionally, online communities are constantly changing. How people use the platform, the types of questions they ask, and even the platform itself can evolve over time, potentially impacting the relevance of our results in the future.

**Internal Validity:** Threats to internal validity can influence the cause-and-effect relationships observed in our study. Measurement bias can arise from the performance of OCR and Image Captioning technologies. The quality and variability of the images within the dataset can affect these technologies' accuracy. The specific implementations we used for OCR and Image Captioning could also introduce bias in our measurements. Another potential source of internal bias is annotation bias. Our study relies on the assumption that identified duplicate questions are accurately labeled. However, the subjective nature of duplicate question identification can introduce bias during the annotation process.

**Construct Validity:** Construct validity refers to the degree to which our study truly captures the concepts we intended to measure. In this case, a potential threat to construct validity is the difficulty of replicating our results. Variations in hardware configurations and computational resources can hinder the reproducibility of our experiments. For this study, we used a specific hardware setup (Apple M2 Pro with 10 cores) and software environment (Python notebooks) to conduct our experiments. If researchers attempt to replicate our findings using different hardware or software, they may encounter discrepancies.

**Conclusion Validity:** It is important to consider how well our conclusions match the actual results of the study. We looked at how well using image data (extracted through OCR and image captioning) helped identify duplicate questions. While there was a slight improvement in accuracy, suggesting that images can provide additional information to the text in question on Q&A platforms, the time and computer power needed for these techniques were significant. Therefore, our conclusion that image features only offer a small benefit while requiring a lot of resources is well-supported by the data we collected and analyzed.

## 9 CONCLUSION

In the dynamic landscape of developer communities, where Q/A platforms are integral to knowledge sharing, our study responds to the evolving communication dynamics by exploring image-based techniques for duplicate question identification on platforms such as Stack Overflow. Despite the prevalent use of images, contemporary duplicate question detection methodologies predominantly rely on text-based analysis.

We investigated techniques to understand text in images (OCR) and describe image content (captions) for duplicate question detection. We also looked into the possible combination of OCR and image captions to make the model more robust and capture more details from the images. These methods yielded a modest improvement (around 1-2%), suggesting images complement the textual content of questions. However, they increased system complexity and processing power demands. Given the minimal improvement versus the significant drawbacks, image features seem to offer limited practical value for duplicate detection tasks and hence, we present it as negative results. Our approach is easy to use and test by others, which paves the way for future research on better ways to find duplicate questions that consider both text and images on platforms like Stack Overflow.

## 10 REPLICATION PACKAGE

For the reproducibility of this study and open science, we provide the community with all the artifacts used in our study. In particular, we provided the source code of our model with documentation and our labeled and unlabelled datasets. The project including all artifacts is available at this google drive link